\documentclass{PoS}
\usepackage{amsmath}
\usepackage[numbers,sort&compress]{natbib}
\usepackage{slashed} 
\newcommand{\avg}[1]{\left\langle #1 \right\rangle} 
\newcommand{\mc}[1]{\mathcal{#1}} 

\title{Towards Partial Compositeness on the Lattice: Baryons with Fermions in Multiple Representations}

\ShortTitle{Baryons with Fermions in Multiple Representations}

\author{
T.A. DeGrand,$^a$
D. Hackett,$^a$ 
\speaker{W.I.~Jay},$^a$
E.T. Neil,$^{ab}$
Y. Shamir,$^c$
B. Svetitsky$^c$ \\
\llap{$^a$} Department of Physics, University of Colorado, Boulder, CO 80309, USA \\
\llap{$^b$} RIKEN-BNL Research Center, Brookhaven National Laboratory, Upton, NY 11973, USA \\
\llap{$^c$} Raymond and Beverly Sackler School of Physics and Astronomy, Tel Aviv University, 69978 Tel Aviv, Israel \\
}

\abstract{We describe our recent lattice study of SU(4) gauge theory with fermions in the fundamental and sextet representations. In this theory, a new type of baryon consists of quarks in both representations. The spectrum of these ``chimera baryons'' has a straightforward interpretation in terms of a non-relativistic quark model based on SU(4). Our results are particularly relevant to composite Higgs models in which the top quark is partially composite.}

\FullConference{34th annual International Symposium on Lattice Field Theory\\
                 24-30 July 2016\\
                 University of Southampton, UK}

\begin{document}

\section{Introduction}

We report results of an exploratory lattice study of the spectrum of SU(4) gauge theory with fermions in two different representations (or ``mixed-rep theories'' for short). Mixed-rep theories are interesting from several perspectives. First, for many years people have speculated about the phase diagram of mixed-rep theories. For example, a mixed-rep system might restore chiral symmetry at a different scale from the deconfinement transition. Another possibility is that chiral symmetries associated with each representation might break at different scales.

Second, mixed-rep theories have appeared in recent models of physics beyond the Standard Model (BSM) in which the Higgs particle arises as a pseudo Nambu-Goldstone boson in a new strongly coupled sector~\citep{Ferretti_SU4_model:2014,Ferretti_partial_compositeness:2013,Ferretti:2016upr}. A new and generic feature of mixed-rep theories is the presence of baryons containing fermions in multiple representations---states we christen ``chimera baryons.'' While their novelty makes chimera baryons interesting in their own right, their chief value for phenomenology lies in composite Higgs models where they appear as top partners---that is, as composite states that mix linearly with the top quark, an idea originally due to Kaplan~\citep{Kaplan:1991dc}.

\section{SU(4) gauge theory with fermions in multiple representations}\label{sec:thy_intro}
Regardless of the fermion representation, gauge invariance prescribes the coupling between fermions and the gauge field. In the fundamental representation, the Dirac term contains the well-known quark-gluon coupling of the form $\bar q^a \slashed{A}^i (T_i)_a^b q_b$, where $T_i$ are generators of the fundamental representation of the gauge group, in our case SU(4). For the sextet---or two-index antisymmetric representation---of SU(4) we can immediately write an analogous expression in terms of the sextet generators $(t_i)_{ab}^{cd}$: $\bar{Q}^{ab} \slashed{A}^i (t_i)_{ab}^{cd} Q_{cd}$. The sextet generators are related to the usual generators through~\citep{Cvitanovic:1976am}:
\begin{align}
(t_i)_{ab}^{cd} = \frac{1}{2} 
	\left[
		(T_i)_a^c \delta_b^d  - (T_i)_b^c \delta_a^d + \delta_a^c (T_i)_b^d - \delta_b^c (T_i)_a^d
	\right].
\end{align}

A notable feature of the sextet representation of SU(4) is that it is a \emph{real} representation. In fact, sextet SU(4) is isomorphic to the fundamental representation of SO(6). In physical terms, quarks and antiquarks are the same in real representations, and so no distinction exists between sextet mesons and diquarks~\citep{DeGrand:2015lna}. Details related to our lattice implementation of this theory are given below in Sec.~\ref{sec:lattice_details}.

In this SU(4) theory, chimera baryons have the schematic form $Qqq$, where $Q$ is a sextet and $q$ is a fundamental fermion.  As always, the Pauli principle must be satisfied for physical states. For chimera baryons this condition implies that the overall wavefunction must be antisymmetric under exchange of the fundamental quarks $q$. We contract the gauge quantum numbers against the Levi-Civita tensor of SU(4) in order construct a singlet: $\epsilon^{abcd} Q_{ab}q_c q_d$. The remaining quantum numbers for the two fundamental fermions are flavor SU(2) and Dirac spin. With the assumption that the spatial wavefunction is exchange symmetric, the situation is then analogous to the $S=-1$ hyperons, with the sextet fermion $Q$ playing the role of the strange quark. For this reason, we simply refer to the three possible chimera states using the names of their QCD counterparts: $\Sigma^*, \Sigma$, and $\Lambda$. Recall that the $\Sigma^*$ is a spin-$3/2$ isotriplet ($I=1$), the $\Sigma$ is a spin-$1/2$ isotriplet ($I=1$), and the $\Lambda$ is a spin-1/2 isosinglet ($I=0$). The quark model predicts the following mass spectrum for these hyperons:
\begin{align}
m_\text{B} &= 
	m_Q +2 m_q  + \frac{\text{const.}}{m_q^2} 
	\left[
		\vec{S}_1 \cdot \vec{S}_2 + \frac{m_q}{m_Q} \vec{S}_Q \cdot (\vec{S}_1 + \vec{S}_2)
	\right] \\
	&\stackrel{QCD}{=} m_Q +2 m_q + \frac{\text{const.}}{m_q^2} \begin{cases}
	+\frac{1}{4} + \frac{m_q}{2m_Q}, \text{ for } \Sigma^*\\
	+\frac{1}{4} - \frac{m_q}{m_Q},  \text{ for } \Sigma \\
	-\frac{3}{4}, \text{ for } \Lambda,
	\end{cases} \label{eq:QCD_quark_model} \\
	&\stackrel{SU(4)}{=} m_Q +2 m_q + \frac{\text{const.}}{m_q^2} \begin{cases}
	+\frac{1}{4} + \frac{m_q}{m_Q}, \text{ for } \Sigma^* \text{ analog} \\
	+\frac{1}{4} - 2\frac{m_q}{m_Q}, \text{ for } \Sigma \text{ analog}\\
	-\frac{3}{4}, \text{ for } \Lambda \text{ analog}.
	\end{cases} \label{eq:su4_quark_model}
\end{align}
The masses $m_q$ and $m_Q$ above are constituent quark masses of the quark model and account for the bulk of the masses of the baryons. In the quark model, a ``color hyperfine interaction'' produces the small splittings above. As with the familiar hyperfine interaction of atomic physics, the color hyperfine interaction goes as the square of the dipole moments, which accounts for the overall $1/m_q^2$ scaling of the splittings. 

The different coefficients for $m_q/m_Q$ in Eq.~(\ref{eq:QCD_quark_model}) (for QCD with fundamental fermions) and Eq.~(\ref{eq:su4_quark_model}) (for mixed-rep SU(4)) arise physically because the sextet quark has two color indices and so feels the color force twice as keenly. More formally, the factors of two come from the ratio of quadratic Casimirs in the sextet and fundamental representations of SU(4).

\section{Lattice details}\label{sec:lattice_details}
\subsection{Simulation details} \label{ssec:simulation_details}
Since this work is an exploratory study, we used existing gauge configurations from a previous lattice investigation of SU(4)~\citep{DeGrand:2016pur}. The configurations from that study were generated with two flavors of dynamical Wilson-clover fermions in the fundamental representation. 

For this study, the necessary valence sextet propagators were cast upon configurations with a volume of $V = 16^3 \times 32$, a gauge coupling $\beta = 10.2$, and a hopping parameter for the fundamental fermions $\kappa_4= 0.1265$.\footnote{The fundamental representation of SU(4) is 4-dimensional, hence the subscript.} This ensemble has fairly heavy quarks, with a pseudoscalar-to-vector mass ratio of $m_\text{PS} / m_\text{V} = 0.337(1) / 0.529(3) = 0.637$. The critical value of $\kappa_4$ is $\kappa_{4,c} = 0.1280$~\citep{DeGrand:2016pur}. Since the sextet fermions were not included in the sea of our configurations, the sextet sector of our theory is quenched.

\subsection{Correlation functions for chimera baryons}

The computational building blocks for this study are hadronic two-point correlation functions. Since construction of fundamental baryonic correlators is standard and since purely sextet baryons have been discussed in a recent paper~\citep{DeGrand:2015lna}, we mention only the necessary details for the discussion of the chimera baryons which follows. Fundamental baryons in SU(4) are bosonic bound states of four fermions $qqqq \equiv q^4$, while sextet baryons are bosonic bound states of six fermions $QQQQQQ \equiv Q^6$. While $QQ$ and $QQQQ$ states do exist, the fact that sextet SU(4) is a real representation forces them to coincide with the mesonic $\overline{Q}Q$ and $(\overline{Q}Q)^2$ states. We restrict our discussion below to the $Q^6$ states.

A chimera baryon interpolating field is $\mc{O}_B^\epsilon = \epsilon_{abcd} Q_\alpha^{ab} q_\gamma^c q_\delta^d C^{\alpha\gamma\delta\epsilon}$ where latin indices indicate SU(4) color and greek indices indicate spin. For brevity we suppress flavor SU(2) indices. Both fermions are 4-component, fully-relativistic Dirac spinors, and $C$ is some combination of gamma matrices that projects onto the spin state of the desired baryon. Since $\mc{O}_B^\epsilon$ couples to a fermion, it carries a free spin index. In our treatment of the problem, we find it useful to work with non-relativistic fermions by projecting onto eigenstates of $P_{\pm} = \frac{1}{2}(1 \pm \gamma_4)$. This projection results in spinor objects with two components which we identify with the familiar spin-up and spin-down states of a non-relativistic fermion.

The numerical building blocks of our correlation functions are propagators of the form:
\begin{align}
D^{-1}_q(m|n)^{a,b}_{\alpha,\beta} \equiv \avg{q(m)^a_{\alpha} \bar{q}(n)^b_\beta},
\end{align}
where $m$ and $n$ are points on the lattice. A similar expression holds for the sextet propagator $D^{-1}_Q$. The color indices $a$ and $b$ are as above,  while $\alpha$ and $\beta$ are ``non-relativistic'' spin indices running from 0 (spin-up) to 1 (spin-down). We then consider a two-point function between the points $m$ and $n$ on the lattice:
\begin{align}
	\begin{split}
		\avg{\mc{O}^\lambda_B(m) \overline{\mc{O}}^\zeta_B(n)}
		&= \epsilon_{abcd} \epsilon_{efgh} C^{\alpha\gamma\delta\lambda} C^{\epsilon\phi\eta\zeta} D^{-1}_Q(m|n)^{ab,ef}_{\alpha,\epsilon} \\
		& \phantom{XXXX}\times \left[ D^{-1}_q(m|n)^{c,g}_{\gamma,\phi} D^{-1}_q(m|n)^{d,h}_{\delta,h} - D^{-1}_q(m|n)^{c,h}_{\gamma,\eta} D^{-1}_q(m|n)^{d,g}_{\delta,\phi} \right] \label{eq:chimera_2pt}
	 \end{split}
\end{align}
Note that the expression in square brackets above includes both a direct and an exchange term. This treatment is valid for states like the charged $\Sigma$ or $\Sigma^*$ in QCD which consist of a single light flavor $u$ or $d$. States like the $\Lambda$, $\Sigma_0$, or $\Sigma_0^*$ inherently contain light quarks of two different flavors $u$ and $d$. Since a $u$ cannot be contracted with a $d$, these states have no exchange term. For the $\Sigma$ and $\Sigma^*$, we consider states with $I_z = 1$ where both the direct and exchange term are present.

The spin projectors $C^{\alpha\beta\gamma\lambda}$ are responsible for isolating the correct spin states for the initial and final baryons. In our non-relativistic setting, it is most transparent to regard the spin projectors as encoding Clebsch-Gordan coefficients which couple the spins. 



\subsection{Qualitative expectations} \label{ssec:numerical_considerations}

Eq.~(\ref{eq:chimera_2pt}) contains the final form of the chimera 2-point function $\avg{\mc{O}^\lambda_B(m)\overline{\mc{O}}^\zeta_B(n)}$, neatly packaged in a convenient form for direct numerical implementation. To construct correlation functions $C(t)$ as functions of time only, we sum over spatial slices at fixed values of t and trace over the free spin indices. One anticipates the ground state to dominate $C(t)$, with exponentially small corrections from excited states. As $t$ approaches the temporal size of the lattice $N_t$, the correlation function becomes sensitive to the lattice size, and $C(t)$ experiences an exponential rise from a state ``propagating backward in time.'' As in QCD studies of baryons, the forward and backward propagating states have different masses, and so the correlation functions are \emph{asymmetric}~\citep{Leinweber:2004it}. Physically, this asymmetry amounts to the well-observed fact that the spectrum of a theory with broken chiral symmetry does not exhibit parity doubling.



\section{Results} \label{ssec:lattice_results}

As discussed in Sec.~\ref{ssec:simulation_details}, we report results from a theory with two dynamical flavors of fundamental fermions in the sea and a single quenched valence sextet fermion. We hold the hopping parameter of the fundamental fermions fixed at $\kappa_4 = 0.1265$ (for both the sea and the valence sectors) so that the lattice spacing is unchanged. We explore the effect of varying the (valence) sextet hopping parameter $\kappa_6$.

\begin{figure}[h]
	\centering
	\includegraphics[scale=0.3]{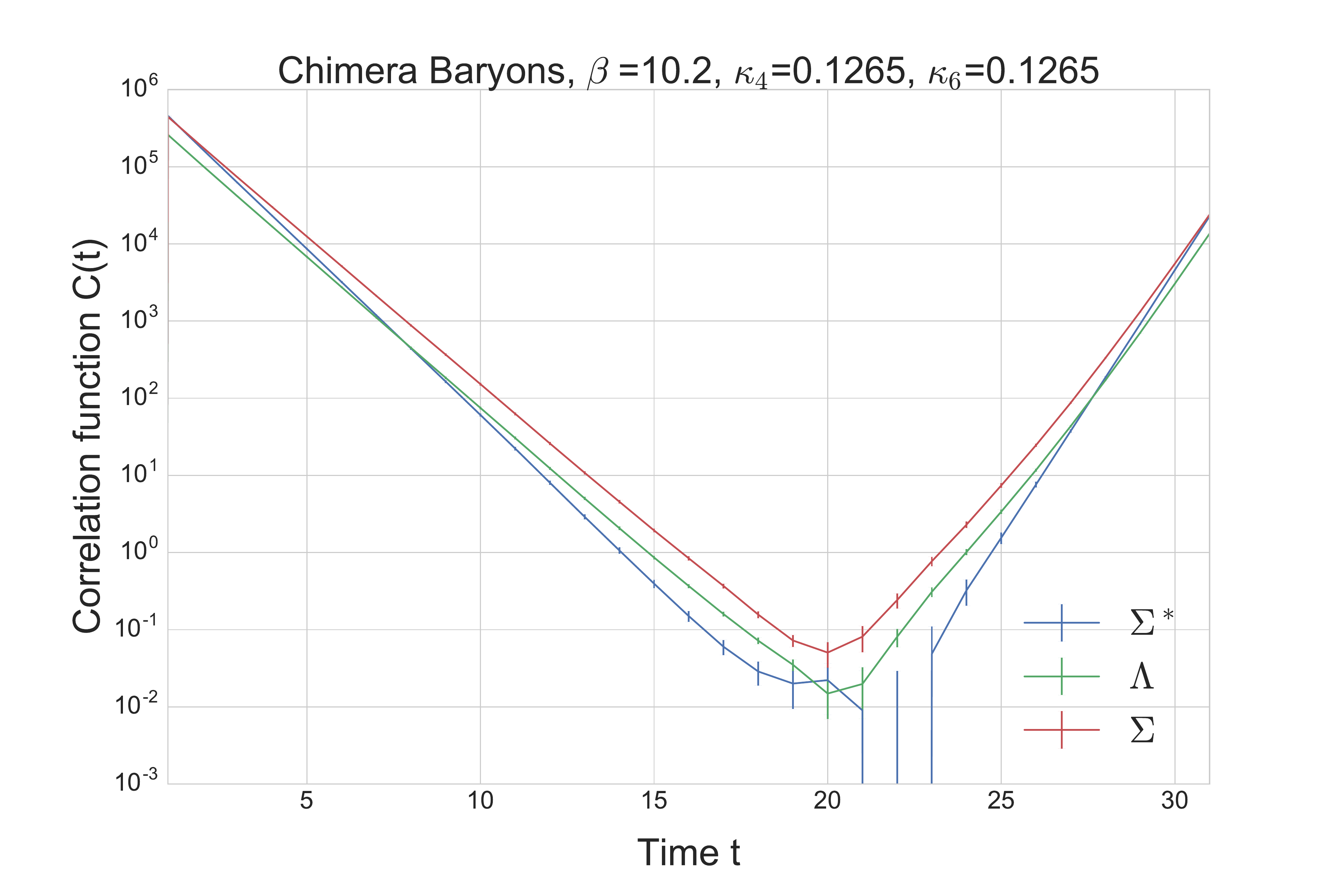}
	\caption{Correlation functions $C(t)$ for the chimera states $\Sigma^*$, $\Sigma$, and $\Lambda$, taken from 50 configurations. Note the anticipated asymmetric shape.}	
	\label{fig:chimera_raw_Ct}
\end{figure}

\begin{figure}[p]
	\centering
	\includegraphics[scale=0.3]{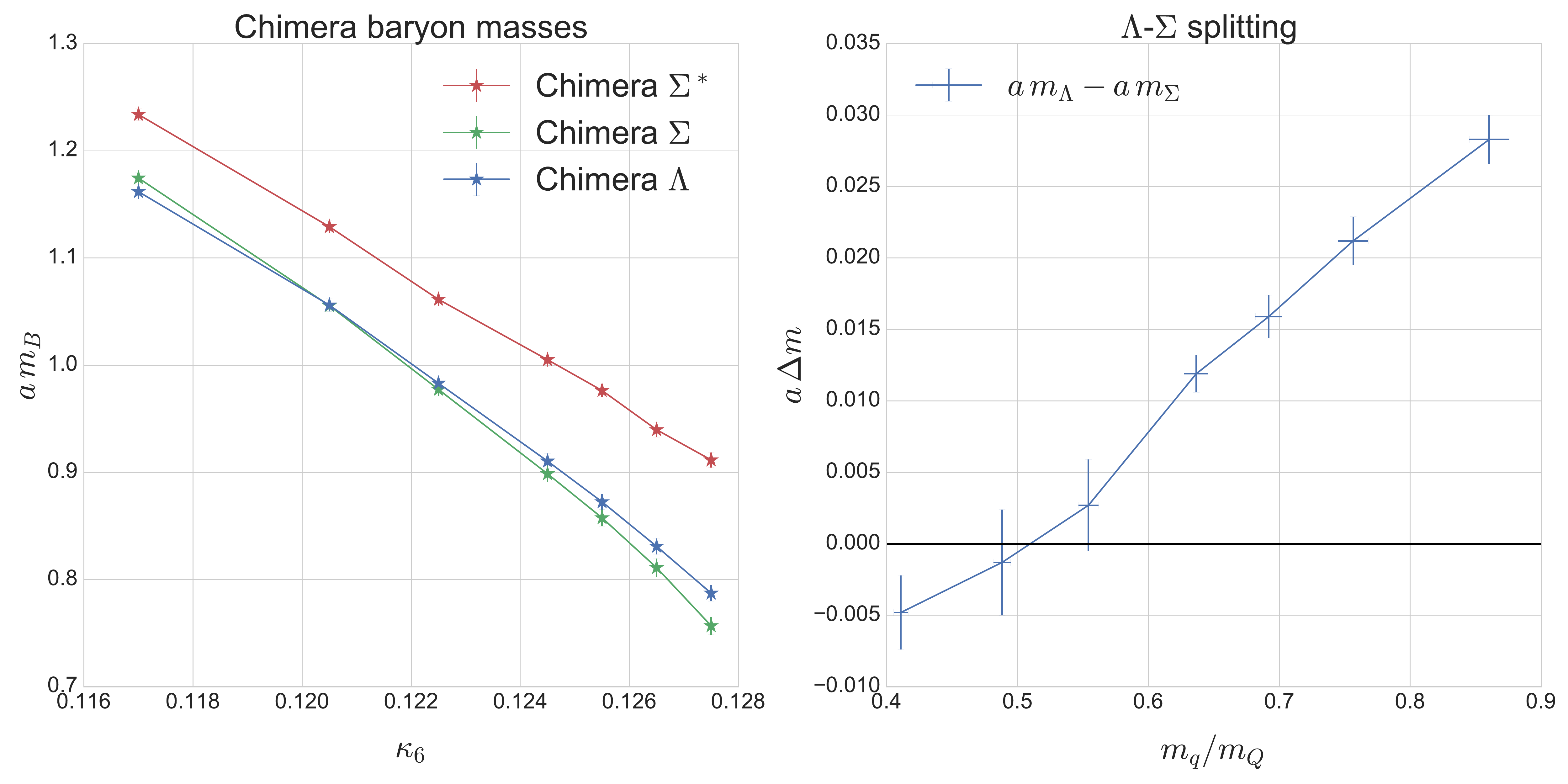}
	\caption{Left panel: the mass spectrum of chimera states in lattice units as $\kappa_6$ is varied with $\kappa_4$ held fixed. Note that the isotriplet $\Sigma$ state becomes the lightest state at large $\kappa_6$, corresponding to light sextet quark mass. Right panel: the mass splitting between the $\Lambda$ and $\Sigma$ states versus the ratio of constituent quark masses $m_q/m_Q$. A positive splitting signals an ``inverted multiplet'' where the $\Sigma$ is the lightest state. Note that the crossing occurs around $m_q/m_Q = 1/2$, as predicted by the quark model.}	
	\label{fig:chimeras}
	\includegraphics[scale=0.3]{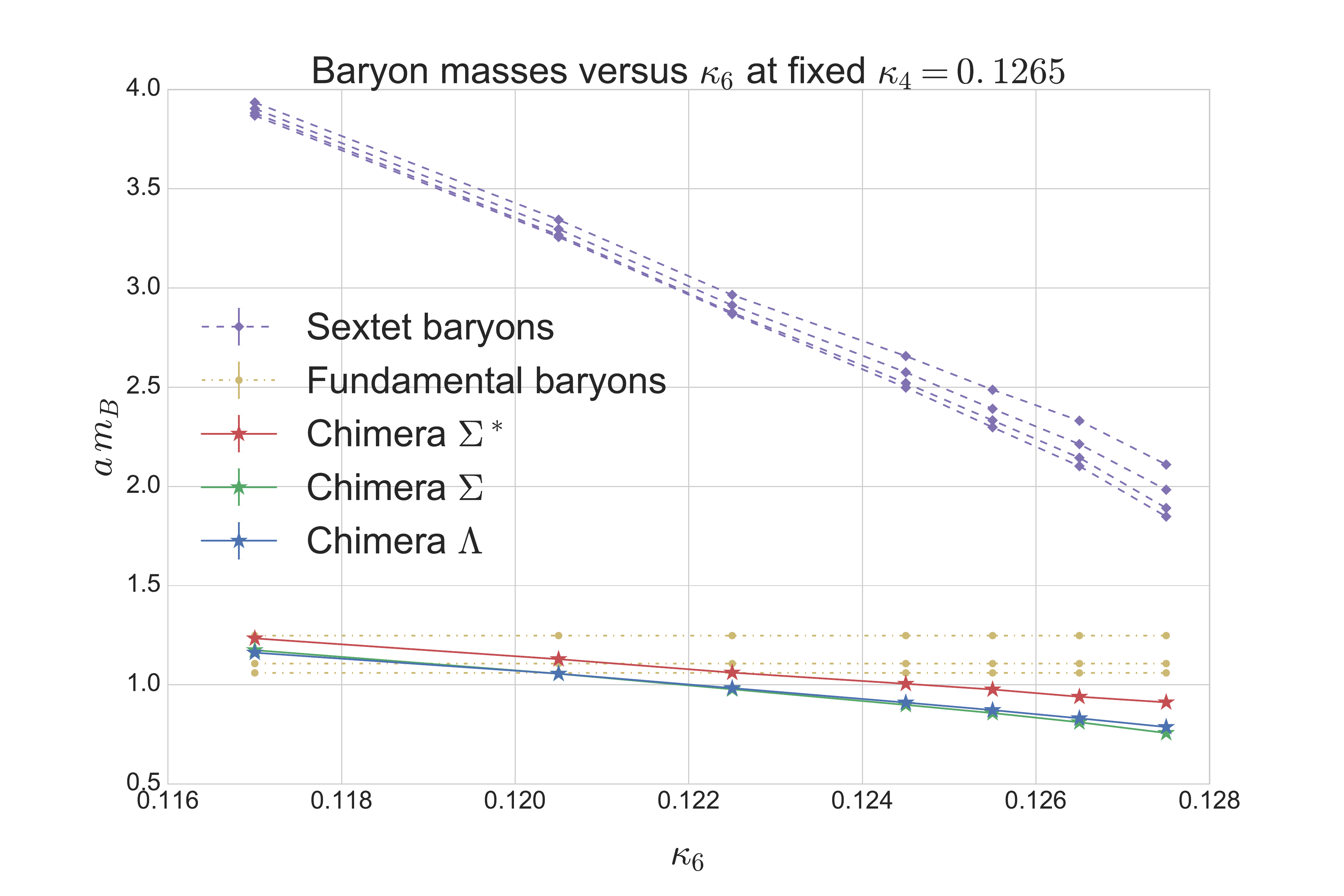}
	\caption{The full spectrum of fundamental, sextet, and chimera baryons in lattice units. Note that at large values of $\kappa_6$---corresponding to light sextet quarks---the chimera baryons are the lightest baryons in the spectrum. Since $\kappa_4$ is held constant, the fundamental baryons have fixed masses.}	
	\label{fig:full_spectrum}
\end{figure}

Figure~\ref{fig:chimera_raw_Ct} shows correlation functions $C(t)$ for the chimera baryons averaged over 50 configurations. These correlators exhibit the advertised asymmetry of Sec.~\ref{ssec:numerical_considerations}. The left panel of Figure~\ref{fig:chimeras} shows the spectrum of chimera baryons as the valence sextet hopping parameter $\kappa_6$ is varied, revealing a crossing between the $\Sigma$  and $\Lambda$ states. At large values of $\kappa_6$---corresponding to small values of the sextet quark mass---the isotriplet $\Sigma$ is the lightest chimera state, in agreement with the quark model prediction of Eq.~(\ref{eq:su4_quark_model}) and different from QCD (where the strange quark has a fixed heavy mass). The right panel of  Figure~\ref{fig:chimeras} shows the behavior of the $(\Lambda-\Sigma)$ mass splitting as a function of $m_q/m_Q$, revealing a crossing very close to the predicted value of 1/2 from Eq.~(\ref{eq:su4_quark_model}). 

In order to construct the right panel of Figure~\ref{fig:chimeras}, it was necessary to estimate the constituent quark masses $m_Q$ and $m_q$. The full spectrum furnishes such estimates for us as follows. Figure~\ref{fig:full_spectrum} shows the entire low-lying baryon spectrum of the theory, with fundamental $q^4$, sextet $Q^6$, and chimera $Qqq$ states. The single-representation objects display rotor splittings consistent with the quark model (and large-N) and analyzed quantitatively in a previous study~\citep{DeGrand:2015lna}. Because of the $J(J+1)$ splittings present in the rotor formula, the color hyperfine interaction vanishes for the lightest ($J=0$) states in the single-representation multiplets. Therefore we can estimate the constituent quark mass $m_Q$ ($m_q$) as 1/6 (1/4) the mass of the lightest state in the $Q^6$ ($q^4$) multiplet.

Finally, Figure~\ref{fig:full_spectrum} reveals that at large values of $\kappa_6$---corresponding to light sextet fermions---the chimera baryons are the lightest baryons in the spectrum. 

\section{Conclusions}\label{sec:conclusions}
We have conducted an exploratory study of the spectrum of SU(4) gauge theory coupled to fermions in two different representations. New fermionic baryons---which we have named chimera baryons---exist in such theories and contain fermions in both representations. We have shown that the baryon spectrum in this theory is qualitatively consistent with expectations from a nonrelativistic quark model based on SU(4), at least in the quenched sextet scenario we considered.

Interestingly for phenomenologists, the chimera states can be the lightest states in the baryon sector, a potentially useful feature for model-builders trying to arrange for a light partner for the top quark. Of course, whether chimera baryons are viable top partners in realistic models also depends on many other questions, such as their location within the full SU(4) spectrum and especially in relation to the Goldstone multiplet(s). We defer further discussion of such points to future work. 

This material is based upon work supported by the U.S. Department of Energy, Office of Science, Office of High Energy Physics, under Award Number DE-SC0010005 (T. D. and E. N.). This work was also supported in part by the Israel Science Foundation under grant no.~449/13. Brookhaven National Laboratory is supported by the U. S. Department of Energy under contract DE-SC0012704.
\bibliographystyle{ssg}
\bibliography{refs}{}

\newcommand{\SortNoop}[1]{}
\begingroup\raggedright\begin{thebibliography}{1}

\bibitem{Ferretti_SU4_model:2014}
G.~Ferretti, {\em {UV Completions of Partial Compositeness: The Case for a
  SU(4) Gauge Group}} {\em JHEP} {\bf 06} (2014) 142,
  \href{http://xxx.lanl.gov/abs/1404.7137}{{\tt 1404.7137}}.

\bibitem{Ferretti_partial_compositeness:2013}
G.~Ferretti and D.~Karateev, {\em {Fermionic UV completions of Composite Higgs
  models}} {\em JHEP} {\bf 03} (2014) 077,
  \href{http://xxx.lanl.gov/abs/1312.5330}{{\tt 1312.5330}}.

\bibitem{Ferretti:2016upr}
G.~Ferretti, {\em {Gauge theories of Partial Compositeness: Scenarios for
  Run-II of the LHC}} {\em JHEP} {\bf 06} (2016) 107,
  \href{http://xxx.lanl.gov/abs/1604.06467}{{\tt 1604.06467}}.

\bibitem{Kaplan:1991dc}
D.~Kaplan, {\em {Flavor at SSC energies: A New mechanism for dynamically
  generated fermion masses}} {\em Nucl. Phys.} {\bf B365} (1991) 259--278.

\bibitem{Cvitanovic:1976am}
P.~Cvitanovic, {\em {Group theory for Feynman diagrams in non-Abelian gauge
  theories}} {\em Phys. Rev.} {\bf D14} (1976) 1536--1553.

\bibitem{DeGrand:2015lna}
T.~DeGrand, Y.~Liu, E.~Neil, Y.~Shamir, and B.~Svetitsky, {\em {Spectroscopy of
  SU(4) gauge theory with two flavors of sextet fermions}} {\em Phys. Rev.}
  {\bf D91} (2015) 114502, \href{http://xxx.lanl.gov/abs/1501.05665}{{\tt
  1501.05665}}.

\bibitem{DeGrand:2016pur}
T.~DeGrand and Y.~Liu, {\em {Lattice study of large $N_c$ QCD}} {\em Phys.
  Rev.} {\bf D94} (2016), no.~3 034506,
  \href{http://xxx.lanl.gov/abs/1606.01277}{{\tt 1606.01277}}.

\bibitem{Leinweber:2004it}
D.~B. Leinweber, W.~Melnitchouk, D.~G. Richards, A.~G. Williams, and J.~M.
  Zanotti, {\em {Baryon spectroscopy in lattice QCD}} {\em Lect. Notes Phys.}
  {\bf 663} (2005) 71--112, \href{http://xxx.lanl.gov/abs/nucl-th/0406032}{{\tt
  nucl-th/0406032}}.

\end{thebibliography}\endgroup

\end{document}